\documentclass{LMCS}

\usepackage[all]{xy}

\usepackage{stmaryrd}
\usepackage{enumerate,hyperref}

\newcommand{\ra}{\rightarrow}
\newcommand{\lra}{\longrightarrow}
\newcommand{\Ra}{\Rightarrow}
\newcommand{\deq}{\stackrel{\scriptscriptstyle{\Delta}}{=}}
\newcommand{\cont}{\leftslice}

\newcommand{\ie}{\textit{i.e.}}

\newcommand{\WHERE}{\mbox{\textbf{where}\;\mbox{}}}
\newcommand{\LET}{\mbox{\textbf{let}\;\mbox{}}}
\newcommand{\IN}{\mbox{}\;\mbox{\textbf{in}\;\;\mbox{}}}

\newcommand{\Ret}{\mathbf{Ret}}
\renewcommand{\Ret}{\mathbf{Wr}}
\newcommand{\Rd}{\mathbf{Rd}}
\renewcommand{\Ret}{\mathbf{Ret}}
\renewcommand{\Rd}{\mathbf{Get}}
\newcommand{\rep}{\mathit{rep}}
\newcommand{\unfold}{\mathit{unfold}}
\newcommand{\out}{\mathit{out}}
\newcommand{\fold}{\mathit{fold}}
\newcommand{\cons}{\mathit{in}}
\newcommand{\eat}{\mathit{eat}}
\newcommand{\teat}{\mathit{eat}}
\newcommand{\hd}{\mathit{hd}}
\newcommand{\tl}{\mathit{tl}}
\newcommand{\Split}{\langle \hd \cdot, \tl \cdot \rangle}

\newcommand{\assoc}{\mathit{assoc}}
\newcommand{\strength}{\mathit{strength}}
\newcommand{\app}{\mathit{app}}
\newcommand{\curry}{\mathit{curry}}

\newcommand{\Set}{\mathsf{Set}}

\newcommand{\BLANK}{\mbox{\_{}}}
\newcommand{\MU}[1]{(\mu\,#1)\,{}}
\newcommand{\NU}[1]{(\nu\,#1)\,{}}

\newcommand{\LAM}[1]{(\lambda \,#1)\,{}}

\newcommand{\SING}[1]{\{\, #1 \,\} }
\newcommand{\SET}[2]{\SING{#1 \, | \, #2}}

\newcommand{\strfby}{\mathop{{\scriptstyle \fatsemi}}}    % {\fatsemi}

\newcommand{\prefixes}[1]{\overline{#1}}

\newcommand{\nil}{{\scriptstyle \diamond}}  %{\langle \rangle}

\newcommand{\COMM}[1]{$\{\mbox{\textit{\small #1}} \}$} 

\newcommand{\K}{\wedge}
\renewcommand{\K}{\mathrel{\mbox{$>\!\!>\!\!=$}}}
\newcommand{\rd}{\mathit{get}}

\usepackage{mathlig}

\def\doi{5 (3:9) 2009}
\lmcsheading%
{\doi}
{1--17}
{}
{}
{Jul.~14, 2008}
{Sep.~15, 2009}
{}   

\begin{document}
\title[Stream Processors using nested fixed pointsshort title]
      {Representations of Stream Processors using nested fixed points}

\author[P.~Hancock]{Peter Hancock\rsuper a}	%required
\address{{\lsuper a}School of Computer Science,
University of Nottingham, Jubilee Campus,
Nottingham, NG8 1BB %, UK
}	%required
\email{hancock@spamcop.net}  %optional
\thanks{{\lsuper a}Hancock's work was supported by EPSERC grant EP/C511964/1.}	%optional

\author[D.~Pattinson]{Dirk Pattinson\rsuper b}	%optional
\address{{\lsuper b}Department of Computing,
Imperial College London,
180 Queen's Gate,
London SW7 2AZ%, UK
}	%optional
\email{dirk@doc.ic.ac.uk}  %optional
%\thanks{thanks 2, optional.}	%optional

\author[N.~Ghani]{Neil Ghani\rsuper c}	%optional
\address{{\lsuper c}Computer and Information Sciences, University of Strathclyde, Livingstone Tower,
26 Richmond St, Glasgow G1 1XH%, UK
}
% School of Computer Science, 
% University of Nottingham, Jubilee Campus,
% Nottingham, NG8 1BB, UK}	%optional
\email{neil.ghani@cis.strath.ac.uk}
       %Nxg@cs.nott.ac.uk}  %optional
%\thanks{thanks 3, optional.}	%optional

%% etc.

%% required for running head on odd and even pages, use suitable
%% abbreviations in case of long titles and many authors:

%% mandatory lists of keywords and classifications:
\keywords{%MANDATORY list of keywords: ?? 
          Streams, continuous functions, initial algebras, final coalgebras.}
\subjclass{%MANDATORY list of acm classifications: ?? 
           68P05, 68N18, 54C35.}
% \titlecomment{OPTIONAL comment concerning the title, \eg, if a variant
% or an extended abstract of the paper has appeared elsewehere}
%%%%%%%%%%%%%%%%%%%%%%%%%%%%%%%%%%%%%%%%%%%%%%%%%%%%%%%%%%%%%%%%%%%%%%%%%%%

%% the abstract has to PRECEED the command \maketitle:
%% be sure not to issue the \maketitle command twice!

\mathlig{|=}{\models}
\mathlig{=>}{\Ra}
\mathlig{->}{\ra}
\mathlig{-->}{\lra}
\mathlig{|->}{\mapsto}
\mathlig{==}{\deq}
\mathlig{<|}{\cont}
%\mathlig{«}{\langle}
%\mathlig{»}{\rangle}

%\maketitle

\begin{abstract}
  We define representations of continuous functions
  on infinite streams of discrete values, both in the case of
  discrete-valued functions, and in the case of stream-valued
  functions.  We define also an operation on the representations of
  two continuous functions between streams that yields a
  representation of their composite.

  In the case of discrete-valued functions, the representatives are
  well-founded (finite-path) trees of a certain kind. The underlying
  idea can be traced back to Brouwer's justification of bar-induction, or
  to Kreisel and Troelstra's elimination of choice-sequences.
  In the case of stream-valued functions, the representatives are
  non-wellfounded trees pieced together in a coinductive fashion from
  well-founded trees.  The definition requires an alternating fixpoint
  construction of some ubiquity.

%   In neither case are the representatives unique.
%   There may be many distinct representatives of extensionally the same
%   function. Nevertheless, the distinctions between them capture
%   genuine differences in computational behaviour.
%   These representations (the data structures and their decoding
%   functions) have a simple (if imperfect) expression in a functional
%   programming language such as Haskell.

%   The ideas extend to functions on final coalgebras
%   for a broad class of functors, though in the general case 
%   to program their representation requires a language with 
%   dependent types.  In some particular cases, only polymorphic
%   recursion is needed.  We hope to publish these results in another paper.

% The representations of functions can be imagined to `eat' their input arguments, under
% the control of a program. Hence the subtitle of the paper, and its occasionally 
% gastronomic 
% nomenclature. 
\end{abstract}

\maketitle

%   The topic of this paper is a method for representing continuous function on 
% infinite objects (such as streams of bits) by data (such as 
% well founded binary trees with values at the leaves). 
% By ``infinite objects'' we mean elements of final
% coalgebras for those functors that can be written as power-series
% (the `normal' functors of Girard \cite{JYG}). Final coalgebras for these functors 
% (at any rate) have a natural topology generalising the Baire topology on streams,
% and this gives rise to an attendant notion of continuous function.

% The idea of the representation can be traced back to Brouwer's bar theorem, 
% and in a sense extends it
% to data types other than streams of discrete data: to potentially infinite terms
% in some finitary unsorted signature.  

\section*{Introduction}

This paper is concerned with the representation and implementation of continuous
functions on spaces of infinite sequences or \emph{streams} of discrete values, such as
binary digits (Cantor space), 
% characters in a finite or countably infinite alphabet,
or natural numbers (Baire space). That is to say, we will look at functions
of type
\[
                f : A^\omega =>X
\]
where $A$ is a discrete space, $A^\omega$ is the space of streams % infinite sequences
of elements of $A$ with the product topology, and $X$ is either
a discrete space $B$, or itself a space of streams $B^\omega$.  We use the symbol
$=>$ for the continuous function space.
Functions of this kind and closely related kinds arise in many contexts in mathematics
and are pervasive in programming, as with pipes, stream input-output and coroutines.

If one is to implement %in any practical sense 
such a function by means
of a program or machine that consumes successive values in an input
stream, and produces a value (all at once in the discrete case, or in
a stream of successive values in the stream-valued case), it seems
necessary that the function be continuous.  Otherwise, the whole input
stream would be needed at once: an output would be forthcoming only
`at the end of time'.  Continuity means that finite information
concerning the output of the function is determined by finite
information concerning its input.  In the simpler, discrete-valued
case, this amounts to the requirement that the value $b = f\,\alpha$
of the function at argument $\alpha$ is determined (or `secured') by some finite
prefix $\overline{\alpha}_n = (\alpha_0, \alpha_1, \alpha_2,
\dots, \alpha_{n-1})$ %(perhaps even the empty prefix) 
of $\alpha$.

It is fairly clear how to represent continuous functions
on $A^\omega$ with discrete values in $B$: take  a well-founded
tree branching over $A$, with $B$'s at the leaves.  Such a tree
represents a continuous function.  Start at the root, then use successive entries in
the argument stream to steer your way along some path to a leaf.  When you
arrive at the leaf, as you inevitably must in view of the tree's being well-founded, 
there is your value for that argument.  We can visualise
the representation as follows.
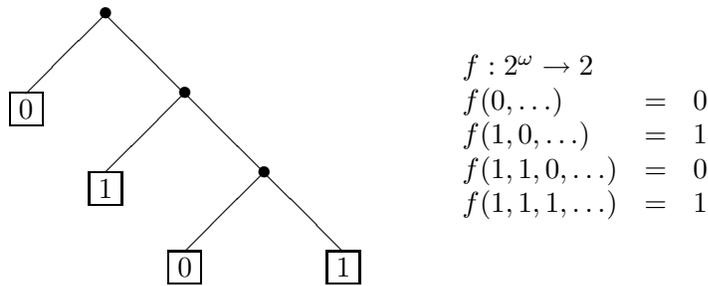
\begin{center}
    \setlength{\unitlength}{3pt}
  \begin{tabular}{cc}
%    \framebox{
    \begin{picture}(40,30)(10,5)
      \put(10,30){\makebox(0,0){$\bullet$}}
      \put(0,20){\line(1,1){10}}
      \put(-2,16){\framebox(4,4)[c]{}}
      \put(0,18){\makebox(0,0){$0$}}
      \put(20,20){\line(-1,1){10}}
      \put(20,20){\makebox(0,0){$\bullet$}}
      \put(8,6){\framebox(4,4)[c]{}}
      \put(10,8){\makebox(0,0){$1$}}
      \put(10,10){\line(1,1){10}}
      \put(30,10){\makebox(0,0){$\bullet$}}
      \put(30,10){\line(-1,1){10}}
      \put(18,-4){\framebox(4,4)[c]{}}
      \put(20,-2){\makebox(0,0){$0$}}
      \put(20,0){\line(1,1){10}}
      \put(38,-4){\framebox(4,4)[c]{}}
      \put(40,-2){\makebox(0,0){$1$}}
      \put(40,0){\line(-1,1){10}}
    \end{picture}
%}
    &  $
    \begin{array}[b]{lcl}
      f : 2^\omega -> 2 \\
      f(0,\ldots) &=& 0 \\
      f(1,0,\ldots) &=& 1 \\
      f(1,1,0,\ldots) &=& 0 \\
      f(1,1,1,\ldots) &=& 1
    \end{array}$ \rule[-1.5cm]{0cm}{3cm}
    \end{tabular}
  \end{center}

At the black inner nodes, the representation `eats' the next entry in the argument stream,
and goes left or right according to whether it's $0$ or $1$; at 
leaves, it `spits' the (boxed) value for that argument.

It should be noted that there will be several (actually, infinitely many) representations of the same function.
For example, the tree below represents the same function as the one above, where (of course) two functions are
the same if their values are equal for all arguments.
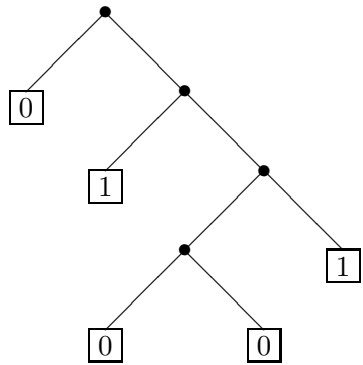
\begin{center}
    \setlength{\unitlength}{3pt}
%  \begin{tabular}{cc}
%    \framebox{
    \begin{picture}(40,50)(10,-15)
      \put(10,30){\makebox(0,0){$\bullet$}}
      \put(0,20){\line(1,1){10}}
      \put(-2,16){\framebox(4,4)[c]{}}
      \put(0,18){\makebox(0,0){$0$}}
      \put(20,20){\line(-1,1){10}}
      \put(20,20){\makebox(0,0){$\bullet$}}
      \put(8,6){\framebox(4,4)[c]{}}
      \put(10,8){\makebox(0,0){$1$}}
      \put(10,10){\line(1,1){10}}
      \put(30,10){\makebox(0,0){$\bullet$}}
      \put(30,10){\line(-1,1){10}}
      %
%      \put(18,-4){\framebox(4,4)[c]{}}
      \put(20,0){\makebox(0,0){$\bullet$}}
      \put(20,0){\line(1,1){10}}
      \put(38,-4){\framebox(4,4)[c]{}}
      \put(40,-2){\makebox(0,0){$1$}}
      \put(40,0){\line(-1,1){10}}
      \put(10,-10){\line(1,1){10}}
      \put(8,-14){\framebox(4,4)[c]{}}
      \put(10,-12){\makebox(0,0){$0$}}
      \put(30,-10){\line(-1,1){10}}
      \put(28,-14){\framebox(4,4)[c]{}}
      \put(30,-12){\makebox(0,0){$0$}}
    \end{picture}
%}
 %    &  $
%     \begin{array}[b]{lcl}
%       f : 2^\omega -> 2 \\
%       f(0,\ldots) &=& 0 \\
%       f(1,0,\ldots) &=& 1 \\
%       f(1,1,0,\ldots) &=& 0 \\
%       f(1,1,1,\ldots) &=& 1
%     \end{array}$ \rule[-1.5cm]{0cm}{3cm}
%     \end{tabular}
  \end{center}

  It is perhaps a little less obvious that the representation sketched
  above is complete: \emph{any} continuous function $f : A^\omega =>
  B$ is representable in this way.  The most straightforward argument
  is irredeemably classical: suppose the function has no
  representation, and derive from this supposition a stream at which
  it is not continuous.  However the completeness of the
  representation can be established constructively, given only the
  validity of a certain principle of called `bar induction', asserting
  the equivalence of two notions of barred-ness, or covering in Baire space.
  Here a `bar' is a monotone subset of $A^\ast$, the set of finite lists of $A$'s. 
  One notion of barred-ness is weak, having the form of a quantification over infinite sequences
  \[ 
  (\forall \alpha : A^\omega)(\exists n : \omega) \,B \langle \alpha_0,\ldots,\alpha_{n-1}\rangle\;.
  \]
  The other notion of barred-ness is strong, being inductively defined, and so having essentially the form of a quantification over subsets of $A^\ast$
  \[
  (\forall U \subseteq A^\ast) B \cup \SET{c : A^\ast}{(\forall a : A)U(c\frown a) } \subseteq U 
  -> U \langle \rangle
  \]
  where $\langle \rangle$ denotes the empty list, and $c \frown a$ 
 the list $c$ with a further entry $a$  at the end.
  Using a variant of this principle, one can show that the 
two notions of continuity on Baire space (one the usual epsilon-delta definition, the other 
defined inductively) coincide.
% well-foundedness for decidable
%   total orderings $\prec$ on $A$, one weak ($(\forall \alpha : A^\omega)\,
%   (\exists n : \omega)\, \alpha_{n+1} \not \prec \alpha_n$, quantifying
%   over infinite sequences) and one strong (asserting the validity of a
%   certain schema of induction, quantifying over predicates of
%   finite lists $A^\ast$ that descend in $\prec$).
  This principle is closely related to Brouwer's `Bar Theorem', for
  which he presented a fascinating but fallacious\footnote{Brouwer did
    not place any restriction such as monotonicity on $B$. As 
    explained by Dummett \cite[pp 68--75]{DUM} this is definitely an error.}
  argument in three articles.  There is an extensive discussion of
  Brouwer's argument by Dummett in \cite[pp 68--75]{DUM}, and a more
  formal analysis of bar induction by Howard and Kreisel in
  \cite{HowardK66}.  A very penetrating discussion of bar induction
  that is closely related to our representation of continuous
  functions by well-founded trees is given by Tait in \cite{Tait68}.

  The inspiration for our representation of continuous functions with
  discrete codomain was in fact Brouwer's argument for bar-induction,
  that conjures an inductive structure from a proof of a $\Pi^1_1$
  statement $\forall \alpha\, \exists n \ldots$.  This inductive
  content was made explicit by Kreisel and Troelstra \cite[8.4,
  p225]{TVD} in the form of the class `K' of neighbourhood functions
  central to their so-called elimination of choice sequences,
  discussed in \cite[pp 75--81]{DUM} and \cite{Tait68}.  In this
  paper, we put this inductive structure into a datatype.
  In fact the
  paper of Tait's just cited contains (at the bottom of p.195) a
  definition of what amounts to the function $\teat$ in section
  \ref{subsec:defeat} below, differing only in notation.  We claim no
  originality for this insight.

Now what about stream-valued continuous functions on $A^\omega$ with
values in $B^\omega$?  The idea is again quite simple, though as far
as we know, new.  It is also difficult to depict.  What we want is a
\emph{non}-wellfounded tree, branching over $A$, along every path of
which there are infinitely many nodes labelled with an element of $B$.
Start at the root, then use successive entries in the argument stream
to steer your way along some path.  Whenever you arrive at a node labelled
with an element of $B$, as you will inevitably do infinitely often, emit that element as the next entry in the
output stream.  It turns out to be straightforward to express the type
of trees we need as a nested fixed point, in which one forms the final
coalgebra of a functor that is defined using an initial algebra
construction.

% Again, our representations will not be unique, for much the same reason as
% in the discrete-valued case.  
Is this representation complete?  It turns out that every
stream-valued function on streams is representable by a
non-wellfounded tree of the kind we have described, though the
argument is perhaps a little intricate.

% \begin{center}
%     \setlength{\unitlength}{3pt}
%   \begin{tabular}{cc}
% %    \framebox{
%     \begin{picture}(40,30)(10,5)
%       \put(10,30){\makebox(0,0){$\bullet$}}
%       \put(0,20){\line(1,1){10}}
%       %\put(0,18){\circle{4}} 
%       \put(-4,16){\framebox(4,4)[c]{}}
%       \put(0,16){\framebox(4,4)[c]{}}
%       \put(-2,18){\makebox(0,0){$0$}}
%       \put(20,20){\line(-1,1){10}}
%       \put(20,20){\makebox(0,0){$\bullet$}}
% %      \put(10,8){\circle{4}} 
%       \put(6,6){\framebox(4,4)[c]{}}
%       \put(10,6){\framebox(4,4)[c]{}}
%       \put(8,8){\makebox(0,0){$1$}}
%       \put(10,10){\line(1,1){10}}
%       \put(30,10){\makebox(0,0){$\bullet$}}
%       \put(30,10){\line(-1,1){10}}
% %      \put(20,-2){\circle{4}} 
%       \put(16,-4){\framebox(4,4)[c]{}}
%       \put(20,-4){\framebox(4,4)[c]{}}
%       \put(18,-2){\makebox(0,0){$0$}}
%       \put(20,0){\line(1,1){10}}
%       \put(36,-4){\framebox(4,4)[c]{}}
%       \put(40,-4){\framebox(4,4)[c]{}}
% %      \put(40,-2){\circle{4}} 
%       \put(38,-2){\makebox(0,0){$1$}}
%       \put(40,0){\line(-1,1){10}}
%     \end{picture}
% %}
%     &  $
%     \begin{array}[b]{lcl}
%       f : 2^\omega -> 2 \\
%       f(0,\ldots) &=& 0 \\
%       f(1,0,\ldots) &=& 1 \\
%       f(1,1,0,\ldots) &=& 0 \\
%       f(1,1,1,\ldots) &=& 1
%     \end{array}$ \rule[-1.5cm]{0cm}{3cm}
%     \end{tabular}
%   \end{center}

Our main contribution, non-trivially extending the state of the art in
the 1960's, is to formulate a represention of stream processing
components (continuous functions between streams, including their
composition), different from that customary in the logical
literature\footnote{According to this, if $k_f$ represents a function
  $f : \omega^\omega ->B$, then $\phi : \omega^\omega -> B^\omega$ is
  represented by $k_f$ where $f(n \strfby \alpha) = \phi(\alpha,n)$.
  This manoeuver %, which arguably, nay surely, qualifies as a `hack',
  works only when $\alpha$ is a stream of natural numbers, or
  encodable as such.}, that fits better with practical implementation
of stream computation.

The datatype of representations provides a convenient basis for
writing stream processing components in a functional programming
language such as Haskell. Nevertheless, the coding in 
Haskell is not entirely satisfying. The chief advantage of using our data type to 
program stream processing components is that it ensures liveness, through
the use of mixed inductive-coinductive types.  The foundations of Haskell are
located in %domain theory, which is to say in 
%an algebraically compact category
%in which initial and final coalgebras coincide. From that perspective, 
a theory of \emph{partial} functions, and not functions in the
standard mathematical sense.  Totality is something extrinsic, beyond
the scope of the type system.  Our approach guarantees
  that the stream processors are total. So it might be better expressed
in a language for \emph{total} functional programming, as advocated by
Turner \cite{Turner04}, and approximated in systems such as Epigram
and Agda.  This means that evaluation of the constructor form of the
value of a function at an argument in its domain must terminate, in
our opinion something to be striven for in a practical programming
language.

It seems that there are lessons to be learnt from this work for the
design of formalisms and systems for developing dependently typed
programs.  It is not yet entirely clear what facilities for
coinductive definition and reasoning such systems need to provide, and
in what form.  It seems firstly that facilities for
inductive-recursive definition may be needed in connection with
coinductive structures: the neighbourhoods in coinductive types have
an inductive-recursive construction. (Admittedly, this structure does
not become fully evident until we consider more general coinductive
datatypes than streams.) Secondly, inductive and coinductive
definitions are sometimes nested within each other (as in
$\NU{X}\MU{Y}B \times X + Y^A$).  Coding our constructions in current
systems for dependently typed programming has revealed a number of
deficiencies and errors in these systems.  Dealing with recursive
definitions in which induction and coinduction are combined needs
careful analysis, that in our opinion should be based on the universal
properties of initial algebras and final coalgebras.

% Brouwer's principle of bar-induction can be interpreted either weakly or strongly.
% Weakly, it is an analysis of what it means to be given a
% discrete-valued function on choice sequences: according to Brouwer,
% it means to be given a well-founded structure built out of steps
% of 3 kinds, one of which can be eliminated.
% Strongly, it is a 
% `factual' claim about all
% constructively defined discrete-valued functions on choice sequences
% (supposing we have an independent way of characterising such functions). .
% Whether by definition or in the nature of things, the representation comes out as complete for constructively
% defined functions.  Here we prefer to compare representations with
% continuous functions, for which we have a straightforward mathematical
% characterisation, Whether continuity coincides with constructively 
% is a question of philosophical dogmatics about which we have no opinion.

The paper is organised as follows.
\begin{enumerate}[$\bullet$] 
\item {}Section~\ref{sec:preliminaries}: preliminaries.
\item {}Section~\ref{sec:discrete-codomain}: we define the representation of 
  the continuous function space $A^\omega => B$ by the datatype 
  $T_A B = \MU{X}B + X^A$ of wellfounded trees branching over $A$ and terminating in $B$, 
  and show it is complete in the sense that
  each such continuous function has a representative (in fact many) in $T_A B$.
  This part of the paper is in essence fairly well known.
\item {}Section~\ref{sec:stream-codomain}: we define the representation of 
  the continuous function space $A^\omega => B^\omega$ by $P_A B = \NU{Y}T_A(B \times Y)$.
  The main contribution here is the proof of completeness, which is not 
  completely straightforward.  The proof is constructive, given completeness
  in the discrete-valued case.
\item Section~\ref{sec:composition}: we define two representations of composition,
as operators of type $P_B C \times P_A B -> P_A C$, and show their correctness. 
As far as we have been able to discover, this representation is new.
\item {}Section~\ref{sec:conclusion}: in conclusion, we summarise what has been done, point out related work,
  and indicate some directions for further work.
\end{enumerate}
The main definitions of the paper can be transcribed quite simply into Haskell.
A Haskell encoding can be found at
\texttt{http://personal.cis.strath.ac.uk/$\mathtt{\sim}$ng/eating.hs}.

% \hrule
% \vspace{1cm}
% \textbf{(Remove TOC eventually)}
% \tableofcontents

\section{Preliminaries}
\label{sec:preliminaries}

We assume the reader is familiar with the categorical notions of product, coproduct, and
exponential, and standard notations associated with these.  We use `$\cdot$' as infix notation
for composition, with (as usual) postponent at left and preponent at right.
  
% We use angle brackets $\langle \BLANK,\BLANK \rangle$ for pairing, subscripts
% $\BLANK_0$ and $\BLANK_1$ for left and right projection. We use $\INL

\subsection{Streams}

If $A$ is a set, we write $A^\omega$ for the set of countably infinite
streams ($\omega$-sequences) of elements of $A$, and $A^\ast$ for the set of
finite sequences (lists) of elements of $A$.  

We use Greek letters $\alpha$, $\beta$, \ldots as variables over
stream types.  We overload the infix operator $(\strfby)$ (with
section notation) our basic means of constructing both streams and
non-empty lists. Thus if $a : A$, then the following functions prefix
$a$ to streams and to lists.
\[     
\begin{array}{l@{\hspace{1cm}}l}
  (a \strfby) : A^\omega => A^\omega &
  (a \strfby) : A^\ast => A^\ast
\end{array}
\]
We also have the empty list $\nil : A^\ast$. 

As destructors of streams we use $\hd$ and $\tl$. 
\begin{displaymath}
  \begin{array}{l}
    \hd : A^\omega -> A \\
    \tl : A^\omega -> A^\omega
  \end{array}
\end{displaymath}
For all $a : A$ and $ \alpha : A^\omega$ we have 
\begin{displaymath}
  \begin{array}{l}
    \hd(a \strfby \BLANK) = a : A \\
    \tl(\BLANK \strfby \alpha) = \alpha : A^\omega\\
    \alpha = (\hd\, \alpha) \strfby (\tl\, \alpha) : A^\omega
  \end{array}
\end{displaymath}
Here we have for clarity written $\BLANK$ for parts of expressions that need not be named.
The destructors $\hd$ and $\tl$ are used implicitly in pattern-matching definitions.
% (Below, we will reuse the operators $\strfby$, $ \hd$ and $\tl$ in a more general situation.).

We sometimes write $\alpha_0$ for $\hd\, \alpha$, and $\alpha'$ for $\tl \, \alpha$.

We use the function $\prefixes{()} : A^\omega -> (A^\ast)^\omega$ which returns the 
stream of finite prefixes of its argument. It is defined 
% (using
% `list-comprehension' notation\footnote{List comprehension notation is a notation
% for writing lists and streams used in the functional programming language Haskell.}) by
% $\prefixes{\alpha} == \nil \strfby \LC{\alpha_0 \strfby t}{t \leftarrow \prefixes{\alpha'}}$\footnote{This is a coiterative definition, \i.e.{} an unfold. The other definition is a fold over the natural numbers.}.
% Equivalently, 
by $\prefixes{\alpha}(0) = \nil$ and
$\prefixes{\alpha}({n+1}) = \alpha(0) \strfby (\prefixes{\alpha'})({n})$.

Streams are endowed with a topology in which the neighbourhoods are given
by finite sequences $c : A^\ast$.  Each such represents the predicate 
$N\,c = \SET{\alpha}{c = \prefixes{\alpha}(\mathit{len}\,c)}$ 
of streams sharing prefix $c$. We usually suppress the distinction
between $c : A^\ast$ and $N\,c \subseteq A^\omega$.  The relation $\alpha \in N\,c$
can be defined by recursion on list $c$. 
% When the stream entries are bits, the topology generated by this family of basic neighbourhoods
% is Cantor space, and when the stream entries  are natural numbers, we have Baire space. When the entries
% belong to a space of discrete states (\eg{} of an automaton), we have trajectories of states from some
% start position, with a `Baire' topology. 

We use $\Ra$ for the continuous function space. Thus $A^\omega => X$ consists of the
continuous functions from $A^\omega$ to $X$, where $X$ is either
a discrete space $D$, or a space  $D^\omega$ where $D$ is discrete. 
\begin{enumerate}[$\bullet$]

\item 
A discrete valued continuous function $f : A^\omega -> D$ is continuous at $\alpha : A^\omega$
if there is some neighbourhood of $\alpha$ throughout which $f$ is constant. 
In other words, there
exists $n \in \omega$ such that $f$
has the same value throughout the neighbourhood
$\overline{\alpha}\, n$. 
\[    \mathit{image} \,f\, (\overline{\alpha}\, n) = \SING{f\, \alpha}
\]
$A^\omega => D$ consists of functions that are continuous throughout $A^\omega$. 
% In other words, there exists $m : A^\omega -> \omega$ such that
% \[    \mathit{image} f \cdot \overline{\alpha} \cdot m\,\alpha = \SING{f(\alpha)}
% \]

\item
A stream-valued continuous function $f : A^\omega -> B^\omega$
is continuous at $\alpha : A^\omega$ if 
$(\forall n \in \omega)(\exists m \in \omega) f(\overline{\alpha}(m)) \subseteq \overline{f\, \alpha}\, n$, or in other words
to find out a finite amount of information about the value, one need only
provide a finite amount of information about the argument.
% there exists a function $m : \omega -> \omega$ such that
% \[           
% \mathit{image}f \cdot \overline{\alpha} \cdot m  \subseteq \overline{f(\alpha)}
% \]
$A^\omega => B^\omega$ consists of functions that are continuous throughout $A^\omega$. 
% This is to say that there is a function $m : A^\omega -> \omega -> \omega$ such that
% \[           \mathit{image}f \cdot \overline{\alpha} \cdot m\, \alpha \subseteq \overline{f(\alpha)}
% \]
If $m$ does not depend on $\alpha$, the function is uniformly continuous.
Such a function $f$ is \emph{contractive} if it decreases the distance between streams.
Prime examples of contractors are the functions $(a \strfby) : A^\omega => A^\omega$,
indexed by $a : A$.

\end{enumerate}

\subsection{Initial algebras and final coalgebras}
\label{sec:init-algebr-final}

We use $\MU{X}F(X) = \mu F$ and $\NU{X}F(X) = \nu F$ to denote initial and final coalgebras for
an endofunctor $F$, typically an endofunctor on the category of sets.

\textbf{Initial algebras}\;
In general we use $\cons$ for the structure map into the carrier of an
initial algebra.
Thus $\cons : F(\mu\,F) -> \mu(F)$. 
Given an algebra $C, \gamma : F C -> C$, we let $\fold(C;\gamma)$,
or simply $\fold\,\gamma$ to denote the unique % coalgebra 
morphism $\delta : \mu\,F -> C$
such that 
\[
\delta \cdot \cons = \gamma \cdot F\,\delta.
\]
We use $\cons^{-1}$ for the inverse of the structure map, namely $\fold(F \cons)$.

Example: finite sequences $A^\ast \deq \MU{X} 1 + A \times X$.
% $A^\ast$ is a functor covariant in $A$.
% Note that $()^\ast$ is the free monad over the functor $(A \times)$.
We use $\nil$ and $(\strfby)$ as constructors associated with $\BLANK^\ast$, so
\[
\begin{array}{l}
\begin{array}{c}
   \xymatrix@R=0pt{1 \ar[rr]^{\nil} && A^\ast \\
             A \times A^\ast \ar[rr]^{(\strfby)} && A^\ast} 
\end{array} \\
\cons = [ \nil | (\strfby) ] : 1 + A \times A^\ast -> A^\ast
\end{array}
\]

Example:
$T_A B \deq \MU{X} B + X^A$, defined in section \ref{sec:discrete-codomain}. 
The bifunctor $T_A B$ is  covariant in $B$, and contravariant in $A$. 
%(We only use covariance in $B$.)
For fixed $A$, $T_A : \Set -> \Set$ is actually the free
monad over the functor $(\BLANK)^A$ (alias $(A \ra)$, known as the reader monad).  
Intriguingly, our constructions all pivot on the freeness of this monad.
$T_A$ is also known as the tree monad. 
We use $\Ret$ and $\Rd$ for the constructors associated with $T_A$. Thus
\[
\begin{array}{l}
\begin{array}{c}
   \xymatrix@R=0pt{B \ar[rr]^{\Ret} && T_A B \\
             (T_A B)^A \ar[rr]^{\Rd} && T_A B} 
\end{array} \\
\cons = [ \Ret | \Rd ] : B + (T_A B)^A -> T_A B
\end{array}
\]

\textbf{Final coalgebras}\;  In general we use $\out$ for the
structure map from the carrier of a final coalgebra. 
Thus $\out : \nu F -> F(\nu F)$.  Given a
coalgebra $C, \gamma : C -> F C$, we use $\unfold(C;\gamma)$, or
simply $\unfold\, \gamma$ (also called the coiteration of $\gamma$) to
denote the unique coalgebra morphism $\delta : C -> \nu F$ such that
\[
\out \cdot \delta = F\,\delta \cdot \gamma
\]
We use $\out^{-1}$ for the inverse of the structure map, namely
$\unfold(F \out)$. 

Example: streams $A^\omega$. We use $\hd$ and $\tl$ to access components of a stream.
$\out = \langle \hd, \tl \rangle : A^\omega -> A \times A^\omega$, while
$\out^{-1} \langle a, \alpha\rangle = a \strfby \alpha$.
%(For $R \deq \nu(T_A \cdot (B \times))$, $\out(p) : T_A(B \times R)$.)

Example:
$P_A B == \NU{X} T_A (B \times X)$, defined in section \ref{sec:stream-codomain}.
% We will frequently suppress mention of the isomorphism 
% $\out : P_A B \cong T_A (B \times P_A B)$.

% The functor $(B \times)$ is reminiscent of both the state monad $B \mapsto S -> B \times S$ (for fixed $S$),
% and the context comonad $ S \mapsto B \times S$ (for fixed $B$). (Interestingly, if $B$ is a monoid,
% the latter functor is also a monad, known as the writer or accumulator monad.)
% $F \mapsto T_A \cdot F$ seems to take
% monads to monads. 

\section{Discrete codomain}
\label{sec:discrete-codomain}

Recall (from section \ref{sec:init-algebr-final}) that $T_A B \deq \MU{X} B + X^A$. 
In this section we define a function $\teat$ of type $T_A B -> A^\omega -> B \times A^\omega$
that allows us to represent continuous functions in $A^\omega => B$ using elements of 
$T_A B$.  Then we give a non-constructive argument that this representation is complete.

\subsection{Definition of \texorpdfstring{$\teat$}{teat}}
\label{subsec:defeat}

Let $M_A B \deq A^\omega -> (B \times A^\omega)$ be the state monad, with state set $A^\omega$.
(The state is the suffix of the input stream that remains unread.)
%, which is to say the future input).  
The unit and bind (infix $\K$) operators of the state monad are as follows.
\[
\begin{array}{l}
  \eta : B -> M_A B \\
  (\K) : M_A B -> (B -> M_A C) -> M_A C \\
  \eta(b) == \LAM{\alpha} \langle b, \alpha \rangle \\
  (m \K f) == \LAM{\alpha} \LET \langle i, \alpha' \rangle = m \alpha \IN f (i, \alpha')
\end{array}
\]
%Here $K$ is written infixed. 
Note that $M_A$ supports the operation of reading one input:
\[
\begin{array}{l}
  \rd : M_A A \\
  \rd\,\alpha  = \langle \alpha_0, \alpha' \rangle
\end{array}
\]
This function plays an important r{\^o}le below in the guise of
$\langle \hd,\tl\rangle : A^\omega \cong A \times A^\omega$.

The most straightfoward definition of $\teat$ is by structural recursion. 
\[
\begin{array}{ll}
  \teat : T_A B &\ra M_A B \\
  \teat(\Ret\,b) &= \eta\,b \\
  \teat(\Rd\,\phi) &= \rd \K (\teat \cdot \phi) 
\end{array}
\]

\subsection{Completeness}

The following result is in essence well known.
\begin{thm} 
\emph{(Completeness of representation of $A^\omega => B$ by $T_A B$.)}  %\hfill
There is a function $\rep : (A^\omega => B) -> T_A B$ 
such that if $f : A^\omega => B$ then $\pi_0 \cdot (\teat (\rep\;f)) = f$
%\eat \cdot \rep = 1_{A^\omega => B}$. 
\end{thm}

Note that $\rep$ picks a representative for a continuous function from
those that give rise
to extensionally the same function.  When $A$ is infinite, there are
uncountably many such representatives. 

\proof%\hfill
%\begin{proof}
\emph{(Classical)}
Suppose %, for a contradiction, that 
that some function  $f : A^\omega => B$ 
has no representative. We `construct' an argument $\alpha : A^\omega$ at which
$f$ is not continuous.  Thence, if $f$ is continuous at all arguments,
there exists some $r : T_A B$ such that $\eat(r)$ equals $f$.

Starting with $f : A^\omega -> B$, 
we `construct' an infinite sequence of functions without representatives.
In the first
place, for some $a : A$, the function $f \cdot (a \strfby)$ has
no representative.  (Else $f$ itself would have a representation.) 
By a form of the axiom of dependent choices, if
$f : A^\omega -> B$ has no representative, then for some $\alpha :
A^\omega$, none of the functions 
\[
f_0 = f, 
f_1 = f_0 \cdot (\alpha_0 \strfby), 
f_2 = f_1 \cdot (\alpha_1 \strfby) = f \cdot (\alpha_0 \strfby) \cdot (\alpha_1 \strfby), \ldots
\]
have
representatives.  In particular, none of these functions can be constant.  It follows that $f$
is not constant in any neighbourhood of $\alpha$, and so $f$ is not 
continuous at $\alpha$.   \qed

% By contraposition, if $f : A^\omega => B$ is everywhere continuous, then it must have
% a representation.  \qed
%\end{proof}

The structure of this %classical
proof is discussed in Dummett \cite[pp 49--55]{DUM}, 
and Troelstra and van Dalen \cite[8.7, p227]{TVD}.
Of course, there are other proofs that do not make use of constructively illicit
forms of contraposition.  For example, there seem to be proofs that use instead monotone bar-induction,
and are arguably intuitionistically valid.
However as indicated by Tait in \cite[pp 194--196]{Tait68}, the best we can hope to
achieve from a constructive point of view is to find models of suitable systems of
constructive reasoning to which we have adjoined an axiom asserting that a
function of type $A^\omega -> B$ continuous in the weak $\epsilon\!\!-\!\!\delta$ sense 
is always continuous in the strong inductive sense.  It is to be expected
that such a model would refute Church's thesis. 

\section{Stream codomain}
\label{sec:stream-codomain}

Recall  (from section \ref{sec:init-algebr-final}) that $P_A B == \NU{X} T_A (B \times X)$.
The previous section gave a complete representation of discrete valued continuous functions $A^\omega => B$,
where $A$ and $B$ are discrete.  We turn now to stream-valued functions.  First we define a function
$\eat_\infty$ with type $P_A B -> A^\omega => B^\omega$.  We have not been able to find a 
similar representation in the literature.  Then we provide it with a right-inverse $\rep_\infty$.

Of course, $A^\omega => B^\omega$ is isomorphic to $(A^\omega =>B)^\omega$, and so its
elements can be represented by streams of representations of $A^\omega => B$.  However such
a representation would be unusable in practice, as the same input stream would have to be scanned
again and again to produce successive items in the output stream.  

%   We \emph{have} come across the
%   following: if $k_f$ represents a function
%   $f : \omega^\omega ->B$, then $\phi : \omega^\omega -> B^\omega$ is
%   represented by $k_f$ where $f(n \strfby \alpha) = \phi(\alpha,n)$.
%   This manoeuver, which perhaps can be criticised as a `hack', works only when
%   $\alpha$ is a stream of natural numbers, or encodable as such.

\subsection{Definition of \texorpdfstring{$\eat_\infty$}{eat_infinity}}

We define $\eat_\infty$ to be the curried form of a function $e$ of type
$(P_A B) \times A^\omega -> B^\omega$ that is continuous in its
second argument.  Since $B^\omega$ is a final coalgebra, 
to define a function into it, it is enough to
define a coalgebra for $(B \times)$ with carrier $P_A B \times A^\omega$.
\[
\xymatrix{
P_A B \times A^\omega \ar[d]^{\out \times 1} \ar[rr]^{e} % = \app \cdot (\eat_\infty \times 1) } 
% && B^\omega \ar[ddd]^{\langle \hd, \tl \rangle}\\
% T_A(B \times P_A B) \times A^\omega 
% \ar[d]^{\app \cdot (\teat \times 1)} && \\
&& B^\omega \ar[dddd]^{\langle \hd, \tl \rangle}\\
T_A(B \times P_A B) \times A^\omega 
\ar[d]^{\teat \times 1} && \\
M_A(B \times P_A B ) \times A^\omega
\ar[d]^{\app} && \\
(B \times P_A B) \times A^\omega \ar[d]^{\assoc}&& \\
B \times (P_A B \times A^\omega) \ar[rr]^{1 \times e} && B \times B^\omega
}
\]
So 
\[     
\begin{array}{l}
  e : P_A B  \times A^\omega -> B^\omega \\
  e = \unfold (\assoc \cdot \app \cdot ((\teat \cdot \out) \times 1)\ .
\end{array}
\]
Then 
\[
\begin{array}{l}
  \eat_\infty : P_A B -> A^\omega => B^\omega \\
  \eat_\infty  = \curry\,e\ .
\end{array}
\]

A more down-to-earth or humane presentation of the definition of $\eat_\infty$ follows, as it might be written
in a functional programming language.
\newcommand{\data}{\ensuremath{\mathbf{data}}}
\newcommand{\Str}{\ensuremath{\mathit{Str}}}
\newcommand{\ass}{\ensuremath{\mathit{as}}}
\[
\begin{array}[t]{l}
  \data\;\;T\,a\,b\; = \; \Ret\,b \;|\;\Rd (a -> T\,a\,b) \\
  \teat :: T\,a\,b -> \Str \, a -> (b, \Str\,a) \\
  \begin{array}[t]{@{}ll}
  \teat\;(\Ret\,b)\,\ass  &= \; (b,\ass)   \\
  \teat\;(\Rd\,f)\,(a:\ass)  &= \; \teat\,(f\,a)\,\ass 
  \end{array} \\[2em]
  \data\;\;P\,a\,b\; = \; \mathbf{P} (T\, a\, (b, P\,a\,b)) \\
  \eat_\infty :: P\,a\,b -> \Str \, a -> \Str\,b \\
  \eat_\infty\,(\mathbf{P}\,t)\,\ass \;=\;
  \begin{array}[t]{@{}l}
    \LET ((b,p),\ass') = \teat\,t\,\ass \IN \\
    b\,:\,\eat_\infty p\,\ass'
  \end{array}
\end{array}
\]

Remark: this definition generalises effortlessly to the case when the codomain
is an arbitrary final coalgebra for a strong functor $F$ (that is, one equipped with
a suitable natural transformation $\strength : F\,X \times Y -> F(X \times Y)$). 
Let $R \deq \nu (T_A \cdot F)$.
\[
\xymatrix{
R \times A^\omega 
\ar[d]^{\out \times 1} 
\ar[rr]^{e'} % = \app \cdot (\eat_\infty \times 1)} 
&& \nu F \ar[ddd]^{\out}\\
T_A(F R) \times A^\omega 
\ar[d]^{\app \cdot (\teat \times 1)} && \\
(F R) \times A^\omega 
\ar[d]^{\strength}&& \\
F(R \times A^\omega) \ar[rr]^{F(e')} && F(\nu F) 
}
\]
% Then 
% \[
% \begin{array}{l}
%   \eat_\infty : R -> A^\omega -> \nu F \\
%   \app \cdot (\eat_\infty \times 1) = \unfold (\strength \cdot \app \cdot ((\teat \cdot \out) \times 1)
% \end{array}
% \]
Though one can thus represent functions from streams into arbitrary final coalgebras,
it is not clear what a completeness result for this general
representation would be.  Without some serious restriction on the functor $F$
it does not seem possible to conjure up a useful topology on the codomain $\nu F$. In
fact, this is possible for functors that represent a %which represent terms in a % countable
single sorted signature of finite arity operators.  
% One needs to
% know what a basic neighbourhood is in the codomain in order
% to give meaning to the notion of a continuous map. 
We hope to
substantiate this remark in a subsequent publication.

%\newpage
\subsection{Definition of \texorpdfstring{$\rep_\infty$}{rep_infinity}}
\label{sec:definition-rep_infty}

The function $\teat_\infty$ allows us to interpret an element of the datatype
$P_A B$ as a continuous function in $A^\omega => B^\omega$. Now we define
a function $\rep_\infty$ that %allows us to 
picks a representative for any such
continuous function.  In the following subsection, we'll show
that $\rep_\infty$ is right-inverse to $\teat_\infty$. 

As the codomain of  $\rep_\infty$ is to be the carrier of a final coalgebra for the functor
$T_A(B \times \BLANK)$, we define $\rep_\infty$ as
the (unique) coalgebra morphism from a coalgebra for the same functor with
carrier $A^\omega => B^\omega$, namely $\rho \cdot \tau$ in the following diagram. 
\[
\xymatrix{
(A^\omega => B^\omega) \ar[rr]^-{\rep_\infty} \ar[d]_{\tau} && P_A B \ar[dd]^{\out} \\ %{\langle \hd, \tl \rangle} \\
T_A B \times (A^\omega => B^\omega) \ar[d]_{\rho} && \\
T_A(B \times (A^\omega => B^\omega)) \ar[rr]^-{T_A(1 \times \rep_\infty)} && T_A(B \times P_A B) 
}
\]
So $\rep_\infty = \unfold (\rho \cdot \tau )$.
Here
\[
\begin{array}{rcl}
  \tau : (A^\omega => B^\omega)     &\ra& T_A B \times (A^\omega => B^\omega) \\
  \tau\,f &=& \langle \rep(\hd \cdot f) , \tl \cdot f \rangle
\end{array}
\]
% Note that $\tau$ is a composite:
% \[\xymatrix{A^\omega => B^\omega 
%   \ar[d]^{\Split} \\ %\langle \hd \cdot, \tl \cdot \rangle } \\
% (A^\omega => B) \times (A^\omega => B^\omega)
%   \ar[d]^{\rep \times 1}\\
% (T_A B) \times  (A^\omega => B^\omega)}
% \]

The other component $\rho$ of the structure map of our $T_A \cdot (B \times)$-coalgebra is
a fold. (For clarity, we give it a more general type than we need.)  It is in some sense a
`fast-forward' operation.
\[
\begin{array}{rcl}
  \rho : T_A B \times (A^\omega => C) &\ra& T_A(B \times (A^\omega => C)) \\
  \rho \langle \Ret\,b, f \rangle &=& \Ret \langle b, f \rangle \\
  \rho \langle \Rd\,\phi, f \rangle &=& \Rd\,\LAM{a} \rho \langle \phi\,a, f \cdot (a \strfby) \rangle
\end{array}
\]
Remarks: 
% First, 
$\rho$ is actually an isomorphism. 
%(There is an
% obvious `fast-backwards' operation.) 
% Second, it can be reformulated as
% morphism from the tree monad $T_A$ to the monad $B |-> (A^\omega => C)
% -> T_A(B \times (A^\omega => C))$.  This latter monad can be described
% as transformation of $T_A$ by the state monad transformer with state
% $(A^\omega => C)$. 
It does not change the shape of a tree,
but only decorates the data stored at its leaves. So, for example,
$(\pi_0 \cdot \eat)(t,\alpha) = (\pi_0 \cdot \assoc \cdot \eat)(\rho(t,f),\alpha)$
for any $t : T_A B$ and $f : A^\omega => C$.

Although $\rep$ cannot be defined constructively, at least without postulating
some form of bar-induction,  the construction of $\rep_\infty$ from $\rep$ 
is a simple matter of programming. 
% , only that $P_A B$ is
% a weakly final coalgebra. 

\subsection{Completeness of \texorpdfstring{$\eat_\infty$}{eat_infinity}}
Now we want to show that the function $\eat_\infty$ is surjective.  It is enough to
show that $\rep_\infty$ is a right inverse for $\teat_\infty$. 
%Statement of completeness:
\begin{thm} \label{thm:stream-complete} \emph{(Completeness of representation of $A^\omega => B^\omega$ by $P_A B$.)}\\
% \[    (\forall\,f:A^\omega => B^\omega, \alpha : A^\omega)\; 
%        \app (\eat_\infty \times 1) (rep_\infty\;f, \alpha) 
%      = \app(f,\alpha)
% \]
% In other words 
\[(\eat_\infty \cdot \rep_\infty) = 1_{A^\omega =>B^\omega}\;.\]
%$\app \cdot ((\eat_\infty \cdot \rep_\infty)\times 1) = \app$. 
\end{thm}
%%%\begin{proof}
\proof%\hfill
We show that the following relation $R$
is a bisimulation on $B^\omega$, and therefore included in the equality relation.
\[
      R = \{ (f\, \alpha, eat_\infty(rep_\infty\,f,\alpha))
          \,|\,
          \alpha : A^\omega, 
          f : A^\omega => B^\omega
          \}
\]
It is enough to prove that if $f : A^\omega => B^\omega$ and $\alpha : A^\omega$, then
\begin{enumerate}[(1)]
\item $\hd(f\,\alpha) = \hd(\eat_\infty(\rep_\infty\,f,\alpha))$, and
\item $\tl(f\,\alpha) \mathrel{R} \tl(\eat_\infty(\rep_\infty \, f,\alpha))$.
\end{enumerate}
As for (1), %the first,
\[
\begin{array}{rcl}
& &  \hd(\eat_\infty(\rep_\infty\, f,\alpha)) \\
&=&  (\pi_0 \cdot \assoc \cdot \eat)(\out(\rep_\infty\,f),\alpha) \\
&=&  (\pi_0 \cdot \assoc \cdot \eat)(
              (T_A \cdot (B \times))\rep_\infty 
                   \cdot \rho 
                   \cdot (\rep \times 1) 
                   \cdot \Split) f
                                    ,\alpha) \\
&=&  (\pi_0 \cdot \assoc \cdot \eat)(
              (T_A \cdot (B \times))\rep_\infty 
                   \cdot \rho)(\rep(\hd \cdot f),\tl \cdot f)
                                    ,\alpha) \\
&=&  \COMM{$(T_A \cdot (B \times))\rep_\infty \cdot \rho$ doesn't affect shape, or first coordinate of data} \\
%& &  (\pi_0 \cdot \eat)(((\rep \times 1) \cdot \Split)f,\alpha) \\
& &  (\pi_0 \cdot \eat)(\rep(\hd \cdot f),\alpha) \\
&=&  \COMM{Completeness in the discrete-valued case} \\
%& &  (\pi_0 \cdot \eat)(\rep(\hd \cdot f),\alpha) \\
& &  \hd(f\,\alpha) \\
\end{array}
\]
As for (2), %the second, 
we 
start by expanding definitions.
\[
\begin{array}{rcl}
& &  \tl(\eat_\infty(\rep_\infty\,f,\alpha)) \\
% &=&  (%\pi_1 \cdot (B\times)
%       \eat_\infty \cdot \pi_1 \cdot \assoc \cdot \eat)
%      (\out(\rep_\infty(f)),\alpha) \\
% &=&  (%\pi_1 \cdot (B\times)
%       \eat_\infty \cdot \pi_1 \cdot \assoc \cdot \eat)(
%               (T_A \cdot (B \times))\rep_\infty 
%                    \cdot \rho 
%                    \cdot (\rep \times 1) 
%                    \cdot \Split) f
%                                     ,\alpha) \\
&=&  (%\pi_1 \cdot (B\times)
      \eat_\infty \cdot \pi_1 \cdot \assoc \cdot \eat)(
              (T_A \cdot (B \times))\rep_\infty 
                   \cdot \rho)(\rep(\hd \cdot f),\tl \cdot f)
                                    ,\alpha) 
%
% &=&  \COMM{$(T_A \cdot (B \times))\rep_\infty \cdot \rho$ doesn't affect shape, or first coordinate of data} \\
% %& &  (\pi_0 \cdot \eat)(((\rep \times 1) \cdot \Split)f,\alpha) \\
% & &  (\pi_0 \cdot \eat)(\rep(\hd \cdot f),\alpha) \\
% &=&  \COMM{Completeness in the discrete-valued case} \\
% %& &  (\pi_0 \cdot \eat)(\rep(\hd \cdot f),\alpha) \\
% & &  \hd(f(\alpha)) \\
\end{array}
\]
We have to show that 
that for all $f : A^\omega => B^\omega$ and $\alpha : A^\omega$,
\[
\tl(f\,\alpha) \mathrel{R}       
              (\eat_\infty \cdot \pi_1 \cdot \assoc \cdot \eat)(
              (T_A \cdot (B \times))\rep_\infty 
                   \cdot \rho)(\rep(\hd \cdot f),\tl \cdot f)
                                    ,\alpha) \;.
\]
By completeness in the discrete-valued case, it is enough to show
that for all $t \in T_A B$, $f' : A^\omega => B^\omega$ and $\alpha : A^\omega$,
\[
f'(\alpha) \mathrel{R}
(\eat_\infty \cdot \pi_1 \cdot \assoc \cdot \eat)(
              (T_A \cdot (B \times))\rep_\infty 
                   \cdot \rho)(t,f')
                                    ,\alpha)\;.
\]
We argue by induction on the wellfounded structure $t$. 
\begin{enumerate}[$\bullet$]
\item In the base case  that $t$ %$\rep(\hd \cdot f)$ 
has the form $\Ret\, b$, calculation shows that
\[
\begin{array}[t]{rcl}
&&(\eat_\infty \cdot \pi_1 \cdot \assoc \cdot \eat)(
              (T_A \cdot (B \times))\rep_\infty 
                   \cdot \rho)(t,f')
                                    ,\alpha) \\
&=& \eat_\infty(\rep_\infty(f'),\alpha)\;.
\end{array}
%\tl(\eat_\infty(\rep_\infty(f),\alpha)) = \eat_\infty(\rep_\infty(\tl \cdot f),\alpha))\;.
\]
% But $(\tl \cdot f)(\alpha) \mathrel{R}   \eat_\infty(\rep_\infty(\tl \cdot f),\alpha))$, so
% $\tl(f(\alpha))  \mathrel{R} \tl(\eat_\infty(\rep_\infty(f),\alpha))$.
But $(f'\, \alpha) \mathrel{R} \eat_\infty(\rep_\infty(f' %\tl \cdot f
                                                   ),\alpha))$, so we are done with this case.

\item
In the step case  that $t$ %$\rep(\hd \cdot f)$ 
has the form $\Rd\, \phi$, calculation shows that
\[
\begin{array}{rcl}
%&& \tl(\eat_\infty(\rep_\infty(f),\alpha)) \\
&&(\eat_\infty \cdot \pi_1 \cdot \assoc \cdot \eat)(
              (T_A \cdot (B \times))\rep_\infty 
                   \cdot \rho)(t,f')
                                    ,\alpha) \\
&=&
(\eat_\infty \cdot \pi_1 \cdot \assoc \cdot \eat)
((T_A \cdot (B \times))\rep_\infty \cdot \rho)
     (\phi(\alpha_0),    
      %\tl \cdot f'
       f' \cdot (\alpha_0 \strfby)
    ,\alpha')
\;.
\end{array}
\]
But by induction hypothesis,
\[
\begin{array}{rcl}
&& (%\tl \cdot 
    f' \cdot (\alpha_0 \strfby))(\alpha') \\
&\mathrel{R}% =
& (\eat_\infty \cdot \pi_1 \cdot \assoc \cdot \eat)
((T_A \cdot (B \times))\rep_\infty \cdot \rho)
     (\phi(\alpha_0),    
      %\tl \cdot 
      f' \cdot (\alpha_0 \strfby)
    ,\alpha')
\;, 
\end{array}
\]
and moreover $(f' \cdot (\alpha_0 \strfby))(\alpha') = f'(\alpha)$. So we are done with this case too.  \qed

\end{enumerate}

\section{Composition}
\label{sec:composition}

In the previous section we defined a complete representation for continuous functions
in $A^\omega => B^\omega$, using elements of $P_A B$.  As continuous functions are
closed under composition, if $p : P_B C$ represents $\phi : B^\omega => C^\omega$,
and $q : P_A B$ represents $\psi : A^\omega => B^\omega$,
then there's some $r : P_A C$ that represents $\phi \cdot \psi$.  However, the argument
for completeness is less than entirely constructive. Can we
directly program such an $r$ from $p$ and $q$?   Yes!  In fact, in at least
two different ways, one `lazy', or demand driven, and the other `greedy', or data driven. The 
computation is reminiscent of cut-elimination in proof theory, though in
this case the objects that interact with each other %being `cut' together 
are infinite, non-wellfounded
trees, rather than wellfounded derivation trees. 

\subsection{Definition of composition as an operation on representatives}
We define (using coiteration) an operation (`$\otimes$') on representations of stream functions
that represents the composition of those functions, in the sense
\[   \eat_\infty(p \otimes q) = \eat_\infty p \cdot \eat_\infty q 
\]
for $p : P_B C$ and $q : P_A B$.  
% Of course, from a non-constructive point of view,
% such an operation must exist, since our representation is complete.  However, we can
% simply program the representation of a composite from the representations
% of its components.

First, we define a coalgebra $\chi$ for the functor $T_A \cdot (C \times)$.  The carrier will
be the product $S \deq T_B (C \times P_B C) \times T_A (B \times P_A B)$. 
First, we present the defining equations for $\chi$ in
pattern-matching format, as they might be written in a functional program.  (This means that in the
third equation, $t_{bc}$ must have the form $\Rd\,\phi$.)  Then we show how to analyse this code
into nested structural recursions, and so demonstrate that $\chi$ is not just a piece of code,
but actually a function defined by 
universal properties of the functors $T_B$ and $T_A$. 
%(The inner recursion can in fact be expressed using the functoriality of $T_A \cdot (B \times)$.
\[
\begin{array}[t]{l@{}l@{}l@{}lll}
  \chi : & \overbrace{T_B (C \times P_B C)}^{\mbox{\textit{postponent}}} \times 
         & \overbrace{T_A (B \times P_A B) }^{\mbox{\textit{preponent}}}
           &\ra 
           %T_A (C \times (T_B (C \times P_B C) \times  T_A (B \times P_A B) ))
           T_A (C \times S) 
           % T_A (C \times P_B C \times P_A B) 
           \\
  \chi \langle  &\Ret\, \langle c,p_{bc}\rangle &, t_{ab} \rangle 
  &= \Ret\, \langle c,\langle\out\, p_{bc}, t_{ab} \rangle \rangle \\
  \chi \langle  &\Rd\,\phi &, \Ret\,\langle b,p_{ab}\rangle \rangle 
  &= \chi \langle\phi\,b ,\out\,p_{ab} \rangle \\
  \chi \langle  &t_{bc} &, \Rd\,\psi \rangle 
  &= \Rd\,\LAM{a}\chi \langle t_{bc},\psi\,a\rangle
\end{array}
\]
Note that in the second equation, $\chi$ at $\Rd\,\phi$ is defined in
terms of $\chi$ at $\phi\,b$, and hence the postponent `goes down' one step in
the outer structural recursion (though the preponent may `go up',
arbitrarily far).  

It is routine to tease the recursion into the form
of nested structural recursions.  The outer recursion is on the structure of the
postponent $T_B (C \times P_B C)$, with an inner or subordinate recursion on the
structure of the preponent $T_A (B \times P_A B)$.  To write it down, we use
a polymorphic function 
\newcommand{\foldT}{\ensuremath{\mathit{fold}}}
\[
\begin{array}[t]{l}
  \foldT : (B -> C) -> ((A -> C) -> C) -> T_A B -> C \\
  \begin{array}[t]{@{}ll}
  \foldT \,p\,g\,(\Ret\,b) &= p\, b \\
  \foldT \,p\,g\,(\Rd\,\phi) &= g\,(\LAM{a : A}\foldT\,p\,g\,(\phi a)) 
  \end{array}
\end{array}
\]
to express structural recursion over wellfounded trees, or in categorical terms the initiality
of $[\Ret|\Rd]$ among algebras $[p|g] : (B + C^A)->C$. 
The definition of $\chi$ can then be given in the form
\[
\begin{array}[t]{l}
\chi \langle t_{bc},t_{ab} \rangle = \foldT\, p\, g\, t_{bc}\, t_{ab} \\
\WHERE
\begin{array}[t]{l}
p : (C \times P_B C) -> T_A (B \times P_A B) -> T_A (C \times S)  \\
p \, \langle c,p_{bc}\rangle\, t_{ab} = \Ret \langle c, \langle \out \, p_{bc},t_{ab} \rangle \rangle
\\[0.5em]
g : (B -> T_A (B \times P_A B) -> T_A (C \times S)) -> T_A (B \times P_A B) -> T_A (C \times S) \\
g\, f = \foldT (\LAM{\langle b,p_{ab} \rangle} f\, b\, (\out \, p_{ab})) \, \Rd    \;.
\end{array}
\end{array}
\]
Note that the carrier for the algebra of the outer recursion is the function space 
$T_A (B \times P_A B) -> T_A (C \times S)$,
while that for the inner recursion is $T_A(C \times S)$. 
% chiw (tbc,tab) = foldT p1 g1 tbc tab    -- recursion on postponent

% p1 :: (c,P b c) -> T a (b,P a b) -> T c S 
% p1 (c,pbc) tab = Ret (b,(unP pbc,tab))

% g1 :: (b -> T a (b,P a b) -> T c S) -> T a (b,P a b) -> T c S
% g1 f = foldT (\(b,pab)->f b (unP pab)) Get     -- inner fold

%    For example, the second and third
% equations may be replaced by the single equation
% \[
%   \chi \langle  \Rd\,\phi , t_{ab} \rangle \;=\; 
%    \mu (T_A[\LAM{\langle b, p_{ab} \rangle}\chi\langle \phi\,b,\out\,p_{ab} \rangle]\,t_{ab}) 
% \]
% where $T_A[m]$ denotes application of the functor $T_A$ to morphism $m$, and $\mu$ denotes the
% multiplication operator of the monad $T_A$. 
% % Note that -- due to the outer recursion being on the postponent -- the last
% % clause in the above definition is needed only for
% % $t_{bc} = \Rd\,\phi$. 
% We prefer to keep the definition in functional programming style to
% facilitate comparision with data-driven composition, defined below.

In this form of composition, priority is given to the postponent's desire to produce
output.  No input is consumed until both the postponent and preponent
are reading.

$\chi$ gives rise to a composition combinator $\otimes$ as follows.
First, $\unfold\, \chi : S -> P_A C$.  We define $\otimes$ by precomposition
with this unfold.
\[
\begin{array}[t]{l@{}l@{}llll}
  \otimes :  P_B C \times  P_A B  -> P_A C \\
  p \otimes q       \deq (\unfold\, \chi) \langle \out\,p,\out\,q\rangle \\
\end{array}
\]
We call $\otimes$ \emph{lazy} composition, since the internal actions
of the composite are the minimum necessary to respond to demand for
data.

Altenkirch and Swierstra noticed that there is another such coalgebra.
We present its definition first in functional programming style, using
pattern matching; below we show how the equations can be teased into
nested recursions.
\[
\begin{array}[t]{l@{}l@{}l@{}lll}
  \chi' : & T_B (C \times P_B C) \times  & T_A (B \times P_A B) 
           &\ra 
%           T_A (C \times (T_B (C \times P_B C) \times  T_A (B \times P_A B) ))
           T_A (C \times S) 
           % T_A (C \times P_B C \times P_A B) 
           \\ 
  \chi' \langle  &t_{bc} &, \Rd\,\psi \rangle &= \Rd\,\LAM{a}\chi' \langle t_{bc},\psi\,a\rangle \\
  \chi' \langle  &\Rd\,\phi &, \Ret\,\langle b,p_{ab}\rangle \rangle &= \chi' \langle\phi\,b ,\out\,p_{ab} \rangle \\
 \chi' \langle  &\Ret\, \langle c,p_{bc}\rangle &, t_{ab} \rangle &= \Ret\, \langle c,\langle\out\, p_{bc}, t_{ab} \rangle \rangle \\
\end{array}
\]
Because of the top-to-bottom reading of the equations, it is implicit
in the last equation that $t_{ab}$ has the form $\Ret (b, p_{ab})$.
Anthropomorphically, this form of composition gives priority to the
preponent's `greedy' desire to read input.  Whereas with the `lazy'
form, output is produced as soon as the postponent is ready,
regardless of the form of the preponent, in this greedier form of
composition no output is produced until \emph{both} the postponent and
preponent are writing.  We call the composition combinator $\otimes'$
to which $\chi'$ gives rise \textit{greedy} composition, since the
internal actions of the composite are driven by the arrival of data at
the input.

What is the mathematical structure of the code for this form of
composition?  Again, it is definition by nested recursion.  One might
think\footnote{As did we, at first.} that the outer recursion is this
time on the structure of the preponent, and the inner recursion on the
postponent.  In fact, this would not work.  In the crucial middle
clause (in which the two components communicate), the postponent `goes
down' in the structural order, while the preponent may `go up',
arbitrarily far.  A more careful analysis shows that, again, the outer
recursion is on the structure of the postponent, with subordinate
recursions on the structure of the preponent.  In fact there is little
formal difference from our definition of $\chi$ above, except that the
base case of the outer recursion uses another inner recursion rather
than a simple explicit definition.  The local function $p$ then
becomes
\[
\begin{array}[t]{l}
p : (C \times P_B C) -> T_A (B \times P_A B) -> T_A (C \times S)  \\
p \, \langle c,p_{bc}\rangle 
= \foldT\, (\LAM{\langle b, p_{ab} \rangle }
         \Ret \langle c, \langle \out \, p_{bc},\Ret\langle b, p_{ab}\rangle \rangle \rangle)\, \Rd
\end{array}
\]
% p2'' :: (b,P a b) -> T c (a,P c a) -> R a b c
% p2'' bpab = foldT (pp'' bpab) Get             -- inner fold, base case

% pp'' :: (b,P a b) -> (a,P c a) -> R a b c
% pp'' (b,pab) apca = Ret (b,(unP pab, Ret apca))

% At a superficial level, it has to be said that the definition of greedy composition is slightly
% more complicated than its lazier variant.  
Unfortunately we currently have little of substance to say about how these forms of composition are related. 
One might well expect that a pipeline implemented with greedy composition
would be less responsive (\ie{} deliver results later) than one expressed with the lazy form.

\subsection{Correctness of composition}
It remains to prove that the two operations that we defined above really represent composition.
This pivots on the uniqueness property of $\unfold\, \chi$.
Exploiting the similarity of the definitions for $\otimes$ and
$\otimes'$ we can state the following basic lemma that applies to
both. 
For the sake of
readability, the isomorphism $\out : P_A B \cong T_A(B \times P_A B)$
is left implicit.

\begin{lem} \label{lemma:compo-props}
Both composition operators $\copyright \in \lbrace \otimes, \otimes' \rbrace$ satisfy the following laws:
%\begin{multicols}{2}
\begin{enumerate}[\em(1)]
\item  $\Ret (c, p_{bc}) \copyright t_{ab} = \Ret (c, p_{bc} \copyright 
t_{ab})$ (where $t_{ab} = \Ret (b, p_{ab})$ in case $\copyright =
\otimes')$
\item $(\Rd\,  \phi) \copyright \Ret (b, p_{ab}) = \phi\,b
\copyright p_{ab}$
\item $p_{bc} \copyright (\Rd\,   \psi) = \Rd\,  
        (\lambda a.\ p_{bc} \copyright \psi\,a)$
(where $p_{bc} = \Rd\,\phi$ in case $\copyright = \otimes$)
\end{enumerate}
%\end{multicols}
\end{lem}

\proof
%%%%%%%%%\begin{proof}
By unfolding the definitions. Actually, it is the desired
effect of the definitions that we have these properties.   \qed
%%\end{proof}

We now %can now appeal to Rutten's \emph{coinductive proof principle} and
set up a bisimulation that shows that 
$t_{bc} \otimes p_{ab}$
and $t_{bc} \otimes' p_{ab}$
really
represents the composite $\eat_\infty (t_{bc}) \cdot
\eat_\infty(p_{ab})$. Again, the isomorphism 
$P_A B \cong T_A(B \times P_A B)$ is left implicit.

\begin{lem} \label{lemma:props}
%For all $\alpha \in A^\omega$, the relation
\[ R = \lbrace (
  \eat_\infty( p_{bc} \copyright t_{ab},\alpha)
             , \eat_\infty(p_{bc}
             , \eat_\infty(t_{ab},\alpha) )) \,|\,
             \alpha \in A^\omega
   \rbrace
\] is a bisimulation on $C^\omega$ if $\copyright \in \lbrace
\otimes, \otimes' \rbrace$.
\end{lem}

\proof%\hfill
It is enough to prove that
\begin{enumerate}[(1)]
\item
  $\hd(\eat_\infty(p_{bc} \copyright t_{ab},\alpha)) = \hd(
  \eat_\infty(p_{bc},
    \eat_\infty(t_{ab}, \alpha)))$
\item
  $\tl(\eat_\infty(p_{bc} \copyright t_{ab},\alpha)), \tl(
  \eat_\infty(p_{bc} , \eat_\infty(t_{ab}, \alpha))) \in R$
\end{enumerate}
for all 
$p_{bc} \in P_B C$ and and all $t_{ab} \in P_A B$ and all $\alpha
\in A^\omega$.
The proof relies on the following identities, which are readily
derived using Lemma \ref{lemma:compo-props}:

\noindent\emph{Case $p_{bc} = \Ret (c, q_{bc})$.}
\begin{align*}
\eat_\infty(p_{bc} \copyright t_{ab},\alpha)
& =  c \strfby \eat_\infty (q_{bc} \copyright t_{ab}, \alpha) \\
\eat_\infty (p_{bc}, \eat_\infty( t_{ab}, \alpha)) 
& = c \strfby \eat_\infty(q_{bc}, \eat_\infty(t_{ab}, \alpha))
\end{align*}
%Then
%\begin{align*}
%\eat_\infty(p_{bc} \copyright t_{ab},\alpha)
%& = \eat_\infty (\Ret(c, q_{bc} \copyright t_{ab}), \alpha) \\
%& =  c \strfby \eat_\infty (q_{bc} \copyright t_{ab}, \alpha)
%\end{align*}
%and
%\begin{align*}
%\eat_\infty (p_{bc}, \eat_\infty( t_{ab}, \alpha)) \\
%& = \eat_\infty(\Ret(c, q_{bc}) , \eat_\infty(t_{ab}, \alpha)) \\
%& = c \strfby \eat_\infty(q_{bc}, \eat_\infty(t_{ab}, \alpha))
%\end{align*}
%using the definition of $\eat_\infty$ (is there a corresponding
%lemma somewhere?) and Lemma \ref{lemma:props}.
%This shows both claims.

\noindent\emph{Case $p_{bc} = \Rd\, \phi$ and $t_{ab} = \Ret(b, t_{ab})$.}
\begin{align*}
\eat_\infty(p_{bc} \copyright t_{ab},\alpha)
& = \eat_\infty((\phi\,b) \copyright t_{ab},\alpha)  \\
\eat_\infty (p_{bc} , \eat_\infty t_{ab} \alpha) 
& = \eat_\infty (\phi\,b, \eat_\infty(t_{ab},\alpha)).
\end{align*}
%
%We have
%\begin{align*}
%\eat_\infty(p_{bc} \copyright t_{ab},\alpha)
%& = \eat_\infty( (\Rd\, \phi) \copyright \Ret(b, t_{ab}),\alpha) \\
%& = \eat_\infty((\phi\,b) \copyright t_{ab},\alpha) 
%\end{align*}
%and
%\begin{align*}
%\eat_\infty (p_{bc} , \eat_\infty t_{ab} \alpha) 
%& = \eat_\infty (\Rd\, \phi, \eat_\infty( \Ret(b, t_{ab}), \alpha)))  \\
%& = \eat_\infty (\Rd\, \phi, b \strfby \eat_\infty (t_{ab}, \alpha) ) \\
%& = \eat_\infty (\phi\,b, \eat_\infty(t_{ab},\alpha)).
%\end{align*}

\noindent\emph{Case $t_{ab} = \Rd\, \psi$.}
\begin{align*}
\eat_\infty(p_{bc} \copyright t_{ab},\alpha)
& = \eat_\infty (p_{bc} \copyright \psi\,a , \alpha) \\
\eat_\infty (p_{bc},\eat_\infty( t_{ab}, a\strfby\alpha))
&  = \eat_\infty(p_{bc},\eat_\infty( \psi\,a, \alpha)).
\end{align*}
%We have
%\begin{align*}
%\eat_\infty(p_{bc} \copyright t_{ab},\alpha)
%& = \eat_\infty( p_{bc} \copyright \Rd\, \psi, a \strfby \alpha \\
%%& = \eat_\infty (\Rd\, \LAM{a} p_{bc} \copyright \psi(a) , a \strfby \alpha) \\
%& = \eat_\infty (p_{bc} \copyright \psi(a), \alpha)
%\end{align*}
%and
%\begin{align*}
%\eat_\infty (p_{bc},\eat_\infty( t_{ab}, a\strfby\alpha))
%& = \eat_\infty(p_{bc},\eat_\infty (\Rd\, \psi, a\strfby\alpha)) \\
%&  = \eat_\infty(p_{bc},\eat_\infty( \psi(a), \alpha)).
%\end{align*}

\noindent
The claim for $\copyright = \otimes$ now follows by 
nested structural recursion,
the outer induction on the postponent, the inner induction on
the preponent; for $\otimes'$ the nesting is reversed.
\qed
%\end{proof}

\begin{cor}
Both $\otimes$ and $\otimes'$ represent composition, \ie{} for all $p_{bc}$ and all
$t_{ab} \in P_A B$ we have
\[
  \eat_\infty(p_{bc} \copyright t_{ab}) = \eat_\infty t_{bc} \cdot
\eat_\infty p_{ab}.
\]
for $\copyright \in\lbrace \otimes, \otimes' \rbrace$.
\end{cor}

\proof
Immediate from the fact that $R$, defined above, is a bisimulation
and the fact that all bisimulations on a final coalgebra are
contained in the diagonal.
\qed

\section{Conclusion, related work}
\label{sec:conclusion}

We have defined computationally natural representations of continuous functions on streams,
and proved completeness of these representations for the classically
understood notion of continuity.  This involved teasing apart %a careful analysis of
the fixed points involved into those that are initial and those that are final. 
We also defined combinators on representations that represent
the composition of the functions they represent.

We consider the main point of this paper to be i) a representation of
stream processors as trees - this ensures that our stream processors
are total as opposed to the partial functions which exist in the
Haskell function space $A^\omega \rightarrow B^\omega$; ii) a
guarantee that all stream processors can be represented by such trees;
and iii) a demonstration that these trees are well suited to
computation --- this takes the form of an implementation of the
composition of stream processing functions directly on the
representatives themselves.

There may also be advantages of a more technical nature. Very often when a
function is represented by a data structure, such as a wellfounded or infinite tree,
the function is automatically `memoised' -- its values for particular arguments
are recorded in the data, and need not be recomputed if they are needed again.
For example, the representation of functions on finitary inductive types by coinductive trees
(in general, final coalgebras for certain rank 2 functors)
discovered by Hinze \cite{Hin00Mem} and Altenkirch \cite{alti:tlca01a} have this property.
The same phenomenon may occur with our representation of stream functions. 
However their work is concerned with functions on inductive types, as
is natural with initial algebras, whereas
ours is primarily concerned with functions on  coinductive types, which
is in the opposite direction from the universal maps associated with final
coalgebras. 

Our representations are not unique, though different representations of the same
function correspond to computationally different behaviour.  Interesting further work might be
to investigate the equivalence relation between representations corresponding
to (extensional) equality between the represented functions.  The relation
is clearly not decidable, and may be hyperarithmetic or worse (when the
data items consumed and produced are natural numbers). 

Another question that may deserve further study is to understand and compare the
relation between the lazy and greedy forms of composition introduced in section \ref{sec:composition}.
More generally, it may be worth investigating whether there is a real connection between
these forms of composition and superficially similar forms of composition in cut-elimination,
and algorithmic game theory.

The set $A^\omega$ of streams of values in a set $A$ is perhaps the simplest example of
a final coalgebra, namely for the functor $(\times A)$,
a close relative of the set of natural numbers that is an initial algebra for the
functor $(+ 1)$.
Final coalgebras are sets of `infinite' values, that can 
model storage, communication and other evolving devices.
In other work that we hope to publish in due course, 
we have generalised Brouwer's representations so as to cover
continuous functions between structures of other coinductive types than streams,
that is to final coalgebras for a useful class of functors beyond $(A \times)$. 
Broadly the same results can be obtained as for the stream case, though the
generalisation involves more mathematical machinery. 
The mathematical
techniques involve working with indexed families of sets,
using an inductive-recursive definition (of such an indexed
family) in a crucial way.  

It may be possible to extend these
techniques yet further to explore
representations of continuous functions on final coalgebras for finitary \emph{indexed} containers,
that are endofunctors on slice categories. 
Some preliminary investigations suggest that this might be
rather laborious.  On the other hand, it could well be
worthwhile.  Endofunctors of that kind would allow us
to model non-wellfounded \emph{proofs}, and so connect
our work with Mints' continuous cut-elimination \cite{mints78:_finit_inves_of_trans_deriv}, analysed
by Buchholz in \cite{BFPT}.  Another connection that might be
made is with Brotherston and Simpson's non-wellfounded
proof systems in \cite{Brotherston-Simpson:07}.
Yet another is with Niwi\'{n}ski and Walukiewicz's 
infinitary proof trees in \cite{NiwWal97}. 

Stream processing is a very venerable approach to systems design.
Streams were used in a central way in the OS6 operating system of Stoy
and Strachey \cite{OS6}, as well as in commercial operating systems. 
 The Unix piping system, introduced by 
McIlroy, is
stream based, with buffering handled by the system.  In
practical programming, a stream facility is often based on something more complicated
than a mathematical stream (involving perhaps EOF, length, buffering,
bounds, putback, ...).  These more feature-full streams inhabit
coinductive types for more elaborate functors than $(A\times)$, but
they are not substantially different. 

The earliest form of IO in functional programming languages was stream
based \cite{Landin}: a executable program was a (possibly asynchronous)
stream processor.  Experience quickly showed it is easy to make
mistakes in programs using asynchronous interfaces.  Mature
implementations of IO interfaces are therefore based on synchronous
processing, consuming response streams to produce request streams, in
a productive or contractive fashion.  Some early functional operating
systems \cite{FunOS} also used streams (sometimes in a ring) for
communication among system processes.

The programming system
Fudgets \cite{Fudgets} is based on a representation of stream
processors similar to the one in this paper, but without our
separation of final from initial fixed points.
Fudgets are a language for asynchronous stream processing. Various combinators
are available for building up stream processors.  
Implementations of Fudgets with Haskell have been used
to build powerful user interaction (mouse, keyboard, display) interfaces.  
% To understand composition of complicated systems from components, 
% there has been considerable interest in data-flow processing, particularly
% when analogue signals are involved, as in 
% control of physical systems with sensors and actuators. 
% For many reasons, synchronous stream processing is of particular
% interest. This means that a single input causes a single output.
The programming system Yampa \cite{Yampa}, which has been used
to produce code for robots (among other things) uses a synchronous dataflow metaphor,
that is well aligned with classical control theory, with its signal processors
and feedback loops.

It seems obvious that the semantics of feedback loops involves fixpoints,
so it may be natural %(or at least, productive) 
to focus on \emph{contractive} functions, because of
Banach's fixed point theorem (see the references in the paper
\cite{BFPT}).  
This states that contractive functions have unique
fixed points. 
In their paper ``Ensuring streams flow'' \cite{telford97ensuring}
Turner and Telford have analysed a productivity requirement for
ensuring unique solutions of recursion equations.  
Productivity seems to be closely related to contractive functions.
From another perspective, Buchholz has designed a calculus for
writing (recursive) stream processing functions, (and even functions
processing certain not-well-founded trees) which ensures that functions
are contractive where required \cite{BFPT}.
We have not specifically examined the representation of contractive
functions, though they are prominent in the form of the functions
$(\alpha_0 \strfby)$ in our constructions. 
Nor have we yet considered
representations of uniformly continuous functions.

The notion of arrow, introduced to functional programming by Hughes
\cite{HughesArrows} was developed to express compositional
infrastructure in programming generalising that of Kleisli morphisms
for a monad, and crucially interacting with a tensor combinator
according to some reasonable laws. The reference \cite{Arrows}
provides a useful perspective. Abstractly, an arrow is a monoid in a
certain category of bifunctors.  Our stream processors behave quite
well with respect to composition $(\cdot)$, but it is not clear to us
how nicely they play with operators such as $+$, $\times$ and other
multi-input combinators.  It may be that one has to get to grips with
notions of fairness, such as fair merging, in connection with such
combinators.  Another direction for further development is to
investigate combinations of stream processors in which, as in many
applications of stream processing, there are forms of feedback, or
looping.

\paragraph{Acknowledgments}
Our colleagues Altenkirch and Swierstra have in unpublished work
considered the broad topic of modelling impure (effectful) phenomena
such as teletype IO \cite{GordonThesis}, mutable heap variables and
multithreading.  We are grateful to them for interesting conversations
on the topic of stream IO, and in particular for pointing out (and
debugging) the `greedy' form of composition mentioned in section
\ref{sec:composition}.  Their model of teletype IO in
\cite{swierstra-altenkirch:beast}, while close to that expounded in
this paper, does not address productivity and continuity.  Finally we
thank the referees for their close scrutiny of the paper, and many
valuable suggestions.

%%%%%%%%%%%%%%%%%%%%%%%%%

\bibliographystyle{abbrv}
\bibliography{stream-eating-lmcs}
\end{document}